\documentclass[10pt,a4paper,aps,showpacs,showkeys,amsmath,amssymb,pra,floatfix,preprint]{revtex4-1}

\usepackage{amsmath}
\usepackage{amssymb}
\usepackage{slashed}
\usepackage{xcolor}
\usepackage{graphicx}
\usepackage{hyperref}

\usepackage{wasysym}
\usepackage{booktabs}
\usepackage{dcolumn}

\DeclareSymbolFont{rsfs}{U}{rsfs}{m}{n}
\DeclareSymbolFontAlphabet{\mathrsfs}{rsfs}

\newcommand{\red}[1]{ #1 }

\newcommand{\llangle}{\langle \! \langle}
\newcommand{\rrangle}{\rangle \! \rangle}
\newcommand{\srf}{\text{s.r.f.}}

\begin{document}

\title{Non-linear Compton scattering of ultrahigh-intensity laser pulses}
\author{Daniel Seipt}
\email{d.seipt@hzdr.de}
\affiliation{Helmholtz-Zentrum Dresden-Rossendorf, PO Box 51 01 19, 01314 Dresden, Germany}

\author{Burkhard K{\"a}mpfer}
\email{kaempfer@hzdr.de}
\affiliation{Helmholtz-Zentrum Dresden-Rossendorf, PO Box 51 01 19, 01314 Dresden, Germany}
\affiliation{Institut f{\"u}r Theoretische Physik, TU Dresden, D-01062 Dresden, Germany}


\begin{abstract}
We present results for the photon spectrum emitted in non-linear Compton scattering
of pulsed ultra-strong laser fields off relativistic electrons
for intensities up to $a_0\gtrsim 100$ and pulse lengths of a few laser cycles.
At ultrahigh laser intensity, it is appropriate to average over the sub-structures of the differential
photon spectrum. Supplementing this procedure with a stationary phase approximation one can evaluate the total emission probability.
We find the photon yield in pulsed fields to be up to a factor of ten
larger than results obtained from a monochromatic wave calculation.
\end{abstract}

\maketitle

\section{Introduction}

The ongoing progress in laser technology has opened an avenue towards studying
strong-field QED effects in ultra-intense laser fields.
Presently, the strongest available laser systems have a few petawatts, with
a focused peak intensity in the order of $I=10^{22} \, \rm W / cm^2$ \cite{Yanovsky:OptEx2008}.
In the near future, the high-intensity frontier will be pushed forward with ELI \cite{ELI}
exceeding $I=10^{24}\, \rm W/cm^2$, eventually.
A Lorentz- and gauge-invariant dimensionless parameter characterizing the
intensity is given by $a_0^2 = 7.9 \times 10^{-19}  I[{\rm W/cm^2}]  \lambda^2[{\rm \mu m}]$
\cite{Salamin:PhysRep2006} where $I[{\rm W/cm^2}]$ is the laser intensity in $\rm W/cm^2$ and
$\lambda [{\rm \mu m}]$ denotes the laser wavelength in $\rm \mu m$.
Thus, values of $a_0$ of several hundreds can be reached in the near future.

Among other topics, the formation of QED avalanches has been discussed recently \cite{Fedotov:PRL2010,Bell:PRL2008} in ultra-intense laser fields,
where a seed-particle leads to the formation of a cascade by consecutive
photon-emission via non-linear Compton scattering and subsequent pair-production processes.
Here, we focus on the non-linear Compton scattering process in laser fields with $a_0 \gg 1$
with emphasis on finite-pulse envelope effects.
The Compton process has been studied for moderately strong laser pulses $a_0 = \mathcal O(1)$
both within classical electrodynamics as non-linear Thomson scattering
\cite{Krafft:PRL2004,Gao:PRL2004,Hartemann:PRE1996,Seipt:PRA2010}
as well as in quantum electrodynamics
\cite{Narozhnyi:JETP83,BocaFlorescu,Seipt:PRA2011,Mackenroth:PRA2011}.
In $a_0 = \mathcal O (1)$ laser pulses, with a duration of several
cycles of the carrier wave, the non-linear Compton spectrum has interesting structures
with many subpeaks per harmonics,
which may be verified experimentally with present technology
\cite{Seipt:PRA2010,Seipt:PRSTAB2011}.
We present here a method for calculating energy-averaged photon emission spectra for $a_0 \gg 1$
laser pulses.

Our paper is organized as follows: In section \ref{sect.matrix.element} we evaluate the matrix element for non-linear Compton scattering.
In section \ref{sect.probability} we use this matrix element to calculate the emission probability,
presenting our result for the phase-space averaged photon yield.
In section \ref{sect.numerical} we present our numerical results before concluding in section \ref{sect.summary}. In appendix \ref{app.beyond.svea} we comment on the slowly varying envelope
approximation. We derive an exact expression for the non-linear phase
including a carrier envelope phase which has the slowly varying envelope approximation
an a limit. Finally, in appendix \ref{app.spa} we have collected details on the stationary phase approximation, which is at the heart of our approach.

\section{Matrix Element for non-linear Compton scattering}
\label{sect.matrix.element}
\subsection{Basic Relations}

The $S$ matrix element for non-linear Compton scattering
---
where an electron with momentum $p$ emits a single photon with momentum $k'$ and leaves the interaction region with momentum $p'$
---
in the Furry picture is
\begin{eqnarray}
 S_{f i} = -i e \int d^4x \overline{ \Psi}_{p'}(x) \slashed \epsilon' \Psi_p(x) \frac{e^{i k'\cdot x}}{\sqrt{2\omega'}},
\label{eq.def.Sfi}
\end{eqnarray}
where the Volkov state $\Psi_p(x)$ \cite{Volkov:1935} is a solution of the homogeneous Dirac equation in an external classical electromagnetic
plane-wave background field with four-potential $A^\mu(\phi)$,
\begin{eqnarray}
  (i\slashed\partial - e \slashed A(\phi) + m ) \Psi_p(x) = 0.
\end{eqnarray}
The Volkov solution reads
\begin{eqnarray}
 \Psi_p(x) &=&  \exp\{ - i p\cdot x - i f_p(x)\} \left[ 1 + \frac{e}{2k\cdot p} \slashed k \slashed A(x) \right] \frac{u_p}{\sqrt{2p^0}} \label{eq.def.Psi}
\end{eqnarray}
where the non-linear phase function is
\begin{eqnarray}
 f_p(x) &=& \frac{1}{2k\cdot p}\int \limits_{0}^{\phi} d\phi' \big[ 2 e p\cdot A(\phi') - e^2 A^2(\phi') \big].
\end{eqnarray}
In light-cone coordinates, e.g.~$x_\pm = x^0 \pm x^3$, $\mathbf x_\perp = (x^1,x^2)$ and $\mathbf x = (x_+,\mathbf x_\perp)$
for the spatio-temporal coordinates,
the laser four-momentum $k$ has only one component which we choose to be $k_-$.
We also define a special reference frame (\srf) in which the electron is initially at rest, i.e.~it has four-momentum $p = (m,0,0,0)$. Assuming a head-on collision of the electrons with the laser pulse,
the laser frequency in the \srf\ is $\omega = (\sqrt{\gamma_0^2 -1} + \gamma_0)\omega_0$,
where $\omega_0$ ($\gamma_0$) is the laser frequency (Lorentz factor of the electron) in the laboratory system.
For the vector potential $A^\mu$ we employ a transverse plane wave, modified by an envelope function $g$ with pulse length parameter $\tau$,
\begin{eqnarray}
 A^\mu &=& A_0 \, g( \phi /\tau ) \, (\epsilon_1^\mu \cos\xi \cos \phi + \epsilon_2^\mu \sin\xi \sin \phi )
\end{eqnarray}
with linear polarization vectors $\epsilon_i \cdot \epsilon_j = - \delta_{i j}$ and $\epsilon_i\cdot k=0$, $i,j\in (1,2)$, 
$\xi$ denotes the polarization state of the laser~\cite{Seipt:PRA2011}.
A complex circular polarization basis $\epsilon_\pm^\mu = \cos\xi \epsilon_1^\mu \pm i \sin\xi \epsilon_2^\mu$
is suitable for the following considerations.
An infinite monochromatic plane wave is recovered by the limit $g\to 1$.
We require that $g$ is a symmetric function of $\phi$ with $g(0) =1$ and $g(\pm \infty)=0$.
The dimensionless laser amplitude $a_0$ is defined with respect to the peak value of the vector potential as $a_0 = e A_0/m$.

The matrix element (\ref{eq.def.Sfi}) reads in the slowly varying envelope approximation
(see~\cite{Seipt:PRA2011} for details), where terms of $\mathcal O(\tau^{-1})$ are neglected
in the exponent (cf.~Appendix \ref{app.beyond.svea})
\begin{eqnarray}
 S_{fi} &=& (2\pi)^3 \delta^3(\mathbf p' + \mathbf k' - \mathbf p) \frac{2}{k_-} \frac{1}{\sqrt{2p^0 2 p'^0 2 \omega'}} \mathrsfs M,
\label{eq.Sfi} \\
\mathrsfs M	
  &=&  \mathrsfs T^0_0 \mathrsfs A^0_0 + \mathrsfs T^1_1 \mathrsfs A^1_1 + \mathrsfs T^1_{-1} \mathrsfs A^1_{-1}
      + \mathrsfs T^2_0 \mathrsfs A^2_0 + \mathrsfs T^2_2 \mathrsfs A^2_2 + \mathrsfs T^2_{-2} \mathrsfs A^2_{-2}, \label{eq.def.M}
\end{eqnarray}
where we have separated the Dirac structures
\begin{align}
 \mathrsfs T^{ 0}_0 &= \bar u_{p'} \slashed\epsilon' u_p, &
\mathrsfs T^{ 1}_{\pm1} &= \frac{ma_0}{4}\bar u_{p'} \left( \frac{\slashed \epsilon_\pm \slashed k \slashed \epsilon' }{k\cdot p'} 
			      + \frac{\slashed \epsilon' \slashed k \slashed \epsilon_\pm}{k\cdot p}  \right) u_p,\\
  \mathrsfs T^{2}_0 &= \frac{m^2a_0^2}{4 k\cdot p k\cdot p'} (k\cdot \epsilon') \bar u_{p'} \slashed k u_p , &
\mathrsfs T^{ 2}_{\pm2} &= \frac{m^2a_0^2}{16 k\cdot p k\cdot p'} 
	    \bar u_{p'} \slashed \epsilon_\pm \slashed k \slashed\epsilon' \slashed k \slashed \epsilon_\pm u_p, 
\end{align}
from the phase integrals
\begin{eqnarray}
 \mathrsfs A^m_n(s) &=& \intop_{-\infty}^\infty d\phi g^m(\phi) e^{i(s-n)\phi - if(\phi)} , \label{eq.def.Amn}
\end{eqnarray}
with $f(\phi) =  f_p(\phi) - f_{p'}(\phi)$ and dimensionless momentum transfer
\begin{eqnarray}
s = \frac{k'_- + p'_- - p_-}{k_-} 
\stackrel{\srf}{=} \frac{\omega'}{\omega}\frac{1}{1-\frac{\omega'}{m}(1+\cos\theta)},  
\label{eq.def.s}
\end{eqnarray}
where the last equality holds in the special reference frame.
The variable $s$ might be interpreted as a continuous number of absorbed photons, as advocated e.g.~in \cite{Seipt:PRA2011,Ilderton:PRL2011}.
The reasoning is based on the fact that the energy-momentum conservation can be written compactly in the suggestive form
\begin{eqnarray}
p + sk = p'+k'. \label{eq.energy.momentum}  
\end{eqnarray}
Since the frequency $\omega'$ of the perturbatively emitted photon has to be positive, also $s$
has to be positive; $\omega'=0$ would imply $s=0$.
This holds in any reference frame due to Lorentz invariance.

The non-linear function $f_p(\phi)$ in the exponent in (\ref{eq.def.Amn}) can be split into a rapidly oscillating contribution
$\tilde f$ and a slow ponderomotive part $\llangle f \rrangle$, i.e.~$f = \tilde f + \llangle f \rrangle$,
where the latter one is averaged over the fast oscillations of the carrier wave. One finds for these two contributions
\begin{eqnarray}
 \tilde f &=& \alpha g(\phi)  \sin (\phi +  \phi_0) + \frac{\beta \cos 2\xi}{2}  g^2(\phi)  \sin 2\phi, \label{eq.f.tilde.svea}\\
\llangle f \rrangle &=& \beta \int_0^\phi d\phi' g^2(\phi') =: \beta G_2(\phi) \label{eq.f.pond.svea}
\end{eqnarray}
with the coefficients
\begin{align}
 \alpha	   &= \sqrt{\alpha_1^2 + \alpha_2^2} 
	   &\stackrel{\srf}{=}& a_0 s \sin \theta \sqrt{ \cos^2 \xi \cos^2\varphi + \sin^2\xi \sin^2 \varphi },& \\
 \beta    &= \frac{m^2a_0^2}{4} \left( \frac{1}{k\cdot p} - \frac{1}{k\cdot p'} \right ) 
          &\stackrel{\srf}{=}& - \frac{a_0^2}{4}(1+\cos\theta) s
	  ,\label{eq.def.beta} &
\end{align}
where the r.h.s.~equalities hold in the special reference frame, and $\phi_0    = \arctan_2(-\alpha_2,\alpha_1)$.
The angles $\theta$ and $\varphi$ are the polar and azimuthal angle, respectively, of the direction of the perturbatively
emitted photon, measured w.r.t.~the z-axis.
Intermediate definitions to arrive at the given expressions for $\alpha$ and $\beta$ are
$\alpha_1  =   \epsilon_1 \cdot P \cos \xi $,
$ \alpha_2  =   \epsilon_2 \cdot P \sin \xi$ and 
$ P_\mu         = ma_0 \left( { p_\mu}/{k\cdot p} - { p'_\mu}/{k\cdot p'} \right)$.

\subsection{Evaluation of the oscillating part and harmonics}

In order to calculate the matrix element (\ref{eq.Sfi}) the integrations in (\ref{eq.def.Amn}) have to be performed.
This can be done by a direct numerical integration as demonstrated
e.g.~in \cite{BocaFlorescu,Seipt:PRA2011} for certain ranges of parameters.
However, if the exponent is very large, i.e.~for $a_0 \gg 1$, this option is inappropriate and one has
to employ different methods for evaluating (\ref{eq.def.Amn}).

The non-periodic oscillating exponential in (\ref{eq.def.Amn}) can be expanded into a Fourier series over the interval $[\phi-\pi,\phi+\pi]$ with the
$\phi$-dependent coefficients~\cite{Narozhnyi:JETP83} according to
\begin{eqnarray}
e^{-i\tilde f} &=& \sum_{\ell = -\infty}^\infty B_\ell(\phi) e^{-i\ell\phi} , \qquad \qquad
B_\ell(\phi) = {\frac{1}{2\pi}} \intop_{\phi-\pi}^{\phi+\pi} d\phi' e^{i\ell\phi' - i\tilde f(\phi')}.
\end{eqnarray}
In the general case of arbitrary elliptical polarization, the coefficients are two-variable one-parameter Bessel functions~\cite{Korsch:JPA2006}
\begin{eqnarray}
 B_\ell(\phi) 
        &=& J_\ell(\alpha g , \beta g^2 \cos 2\xi \, / 2  ; \phi_0)  
	= \sum_{s=-\infty}^\infty J_{\ell-2s}(\alpha g) J_s(\beta g^2 \cos2\xi\, /2 ) e^{-i(\ell-2s)\phi_0}.
\end{eqnarray}
We focus here on the case of a circularly polarized laser pulse
(i.e.~$\cos2\xi=0$), where the coefficients $B_\ell$ simplify to ordinary
Bessel functions of the first kind multiplied by a phase factor $B_\ell(\phi) = e^{-i\ell \phi_0} J_\ell\big(\alpha g(\phi) \big)$.

After Fourier expansion, the matrix element (\ref{eq.def.M}) turns into a sum over partial amplitudes
\begin{eqnarray}
\mathrsfs M &=& \sum_{\ell=-\infty}^\infty \mathrsfs M_\ell \label{eq.M.expansion}, \\
\mathrsfs M_\ell &=&
 e^{-i\ell\phi_0}\Big[
\mathrsfs T^0_0  c_\ell^0(s-\ell) 
+ \mathrsfs T^2_0  c_\ell^2(s-\ell) 
+ e^{i\phi_0}\mathrsfs T^1_1   c_{\ell-1}^1(s-\ell)
+ e^{-i\phi_0}\mathrsfs T^1_{-1}  c_{\ell+1}^1(s-\ell)
\Big] 
\label{eq.M.expanded}
\end{eqnarray}
with purely real coefficients
\begin{eqnarray}
c_{\ell-n}^m(s-\ell) &=&  \intop_{-\infty}^\infty d\phi J_{\ell-n}(\alpha g) g(\phi)^m \exp \{i (s-\ell) \phi - i \beta G_2(\phi) \} \label{eq.def.c}
\end{eqnarray}
with $G_2(\phi) = \int_0^\phi d\phi'g^2(\phi')$ from (\ref{eq.f.pond.svea}).
In the expansion (\ref{eq.M.expansion}), the integer $\ell$ is the net number of absorbed laser photons by the electron
thus labeling the harmonics. The seeming contradiction of continuous $s$ and integer $\ell$ can be resolved easily:
In the energy momentum conservation ($\ref{eq.energy.momentum}$) the photon four-momentum
$k$
is calculated with the central frequency $\omega$.
However, the pulsed laser field has a finite energy bandwidth $\Delta \omega/ \omega \propto 1/\tau$. Therefore, it appears as if there would be a continuous photon number. Also the emitted photon spectrum is a continuous spectrum.
It is not possible to write the energy-momentum conservation as $p + \ell k = p'+k'$.
Consequently, each of the harmonics $\ell$ has a broad support, not only on a delta comb, as for monochromatic bandwidth-free background fields.

As a consequence of energy conservation,
the absorbed photon number has to be positive, $\ell>0$, since otherwise the 
energy $\omega'$ would become negative.
Thus, the sum in (\ref{eq.M.expansion}) starts at $\ell=1$.
The precise location of the support of the harmonics is determined by the regions where
stationary phase points $\phi_\star$ of
(\ref{eq.def.c}) exist on the real axis, i.e.~for real solutions
of the equation
\begin{eqnarray}
g(\phi_\star) = \sqrt{\frac{s-\ell}{\beta}}, \label{eq.stationary.phase}
\end{eqnarray}
which are always found as pairs $\pm \phi_\star$ where we define $\phi_\star > 0$.
In (\ref{eq.stationary.phase}), both the argument of the square root has to be positive and the value of the square root has to be larger than zero and smaller than unity due to the assumptions made for the pulse envelope $g$.
Independent of the explicit shape of the pulse the allowed range of $s-\ell$ is bounded by $\beta \leq s-\ell \leq 0$
(note that $\beta<0$). Outside this region the stationary phase has an imaginary part leading to an exponential suppression of the
coefficient functions $c^m_{\ell-n}(s-\ell)$.
Explicit solutions for $\phi_\star$ are listed in Tab.~\ref{tab.stationary.phase} for
the hyperbolic secant pulse $g(\phi) = 1/\cosh \{\phi /\tau\}$ and the Gaussian pulse $g(\phi) = \exp \{ -\phi^2/2\tau^2\}$.
In particular, in our numeric calculations we use the hyperbolic secant pulse.

When passing from distant past, $\phi = -\infty$, to the distant future, $\phi=+\infty$,
the coefficients pick up a ponderomotive phase shift
\begin{align}
\Delta f =  \lim_{\phi\to\infty}\llangle f\rrangle -  \lim_{\phi\to -\infty}\llangle f\rrangle = h \beta \tau
\label{eq.def.deltaf}
\end{align}
where the proportionality factor $h$ depends on the explicit shape of the pulse (e.g.~$h=2$ for the hyperbolic secant pulse
and $h=\sqrt{\pi}$ for a Gaussian pulse). 
\begin{widetext}
\begin{table}[!t]
 \begin{center}
 \begin{tabular}{lcc} \toprule[1pt]
  & hyperbolic secant & Gaussian  \\ \midrule[0.75pt]
  $g(\phi)$  & $\frac{1}{\cosh \frac{\phi}{\tau}}$ & $ \exp \left\{ \frac{\phi^2}{2\tau^2} \right\}$  \\
  $g'(\phi)$  & $ - \frac{1}{\tau} g \sqrt{1-g^2}$ & $ - \frac{\phi}{\tau^2} g$  \\
  $g''(\phi)$ & $ \frac{g}{\tau^2} (1-2g^2)$ & $ (\frac{\phi^2}{\tau^4} - \frac{1}{\tau^2}) g$  \\
  $g''(0)$  &  $-\frac{1}{\tau^2}$ & $ -\frac{1}{\tau^2}$  \\ 
  $G_2(\phi)$& $\tau \tanh \frac{\phi}{\tau}$ 
  				& $\frac{\sqrt\pi }{2} \tau {\rm erf} \frac{\phi}{\tau}$ \\
  $\phi_\star$  & $ \tau {\rm Arcosh} \sqrt{\frac{\beta}{s-\ell}}$ & $ \tau \sqrt{\ln \frac{\beta}{s-\ell}}$ \\ 
  $g'(\phi_\star)$ & $ -\frac{1}{\tau} \sqrt{1 - \frac{s-\ell}{\beta}} \sqrt{\frac{s-\ell}{\beta}}$ &
    $ - \frac{1}{\tau} \sqrt{ - \frac{s-\ell}{\beta} \ln \frac{s-\ell}{\beta}}$   \\ 
\bottomrule[1pt]
 \end{tabular}
 \end{center}
 \caption{Explicit forms of the pulse shape functions $g$, their derivatives and the ponderomotive integrals $G_2$. Relations for the stationary phase points $\phi_\star$ for these pulse shapes are also provided.}
 \label{tab.stationary.phase}
\end{table}
\end{widetext}

\subsection{Monochromatic limit}

To contrast the realistic case of a finite-duration laser pulse with the idealized case of a infinitely long monochromatic wave we consider the reduction of the former to the latter.
In the limit of a monochromatic plane wave, $g\to 1$, the coefficients (\ref{eq.def.c}) condense to
\begin{eqnarray}
 c_{\ell-n}^m(s-\ell) &\stackrel{g\to1}{\to}&  (2\pi)\delta(s- \ell- \beta) J_{\ell-n}(\alpha) 
\end{eqnarray}
with support on a delta comb at the locations
\begin{eqnarray}
 \omega'_\ell(\theta) &=& \frac{\ell \omega}{1 + \left[ \frac{a_0^2}{4} + \ell \frac{\omega}{m}\right](1+\cos\theta)},
\label{eq.omega.nonlinear}
\end{eqnarray}
coinciding with the lower boundary of the harmonic support in the case of pulsed laser fields.
With this, the $S$ matrix reads
\begin{eqnarray}
 S_{fi} &=& (2\pi)^4 \sum_\ell \delta^4(\tilde p+\ell k - \tilde p'-k')   M_\ell,\\
M_\ell &=&  e^{-i\ell \phi_0} 
\big[
  (\mathrsfs T^0_0  + \mathrsfs T^2_0 ) J_\ell(\alpha) 
  + \mathrsfs T^1_1 e^{i \phi_0}  J_{\ell-1}(\alpha)
  + \mathrsfs T^1_{-1} e^{-i \phi_0}  J_{\ell+1}(\alpha) 
\big],
\end{eqnarray}
which coincides with textbook results (e.g.~\cite{Landau4}) and
where $\tilde p = p + \frac{m^2a_0^2}{4k\cdot p}k$ denotes the intensity dependent quasi-momentum of the initial electron
with the 
effective mass $m_\star^2 = \tilde p \cdot \tilde p = m^2 (1+a_0^2/2)$.
Thus, in a monochromatic wave, the ponderomotive part of the
exponent provides a {constant rate} of phase shift $\llangle f \rrangle \stackrel{g\to1}{\to} \beta \phi$ which leads to the build-up of quasi-momentum and effective mass.
In a pulsed laser field, on the contrary, we find a finite total phase shift $\Delta f$, cf.~(\ref{eq.def.deltaf}).

\section{Photon emission probability}
\label{sect.probability}
The differential photon emission probability (or photon yield) per laser pulse and per electron is defined by (cf.~\cite{Seipt:PRA2011})
\begin{eqnarray}
 \frac{dN}{d\omega' d\Omega} &=& \frac{1}{2} \sum_{\rm spin,pol} \frac{e^2 \omega'}{{64}\pi^3 k\cdot p k\cdot p'} |\mathrsfs M|^2.
\label{eq.def.rate}
\end{eqnarray}
The cross section for a particular process is obtained by dividing the photon emission probability
by the integrated flux of the laser pulse $N_L$, i.e.~$d\sigma = dN/N_L$, where $N_L = \frac{\omega a_0^2 m^2}{2e^2} T_{\rm eff}$ and the effective
interaction time $T_{\rm eff}$ is determined uniquely by the pulse envelope 
\begin{equation}
T_{\rm eff} = \omega^{-1} \int_{-\infty}^\infty d\phi g^2 (\phi). \label{eq.Teff}
\end{equation}
The interaction time is also related to the total ponderomotive phase shift via $\Delta f = \beta \omega T_{\rm eff}$.
In (\ref{eq.def.rate}), the \emph{coherent} sum over all harmonics is a double sum
\begin{eqnarray}
\left| \mathrsfs M \right|^2 &=& 
 \sum_{\ell,\ell'} \mathrsfs M_\ell^* \mathrsfs M_{\ell'} 
=  \sum_\ell  \left| \mathrsfs M_\ell\right|^2 + \sum_{\ell'\neq \ell} \mathrsfs M_\ell^* \mathrsfs M_{\ell'} , \label{eq.rate.doublesum}
\end{eqnarray}
where we separate the \emph{incoherent} diagonal contributions from the off-diagonal elements which lead to interferences between different harmonics
whenever they are overlapping. The overlapping of harmonics can happen only for pulsed fields because of the broad support of each harmonic;
it does not appear for monochromatic laser fields where the support is a delta comb.

\begin{figure}
\includegraphics[width=5cm]{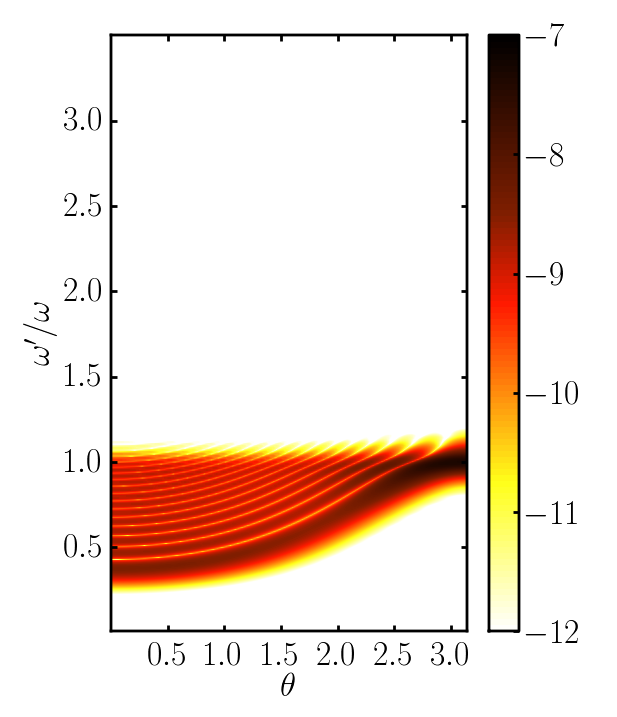}
\includegraphics[width=5cm]{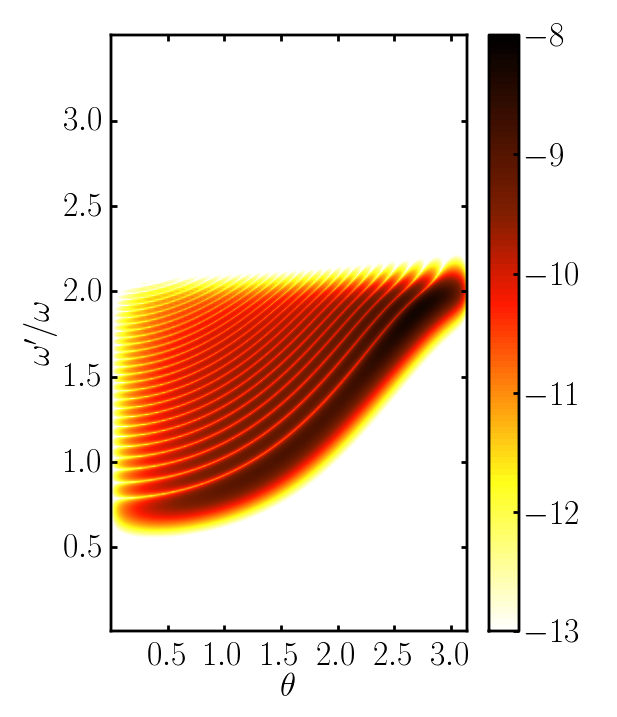}
\includegraphics[width=5cm]{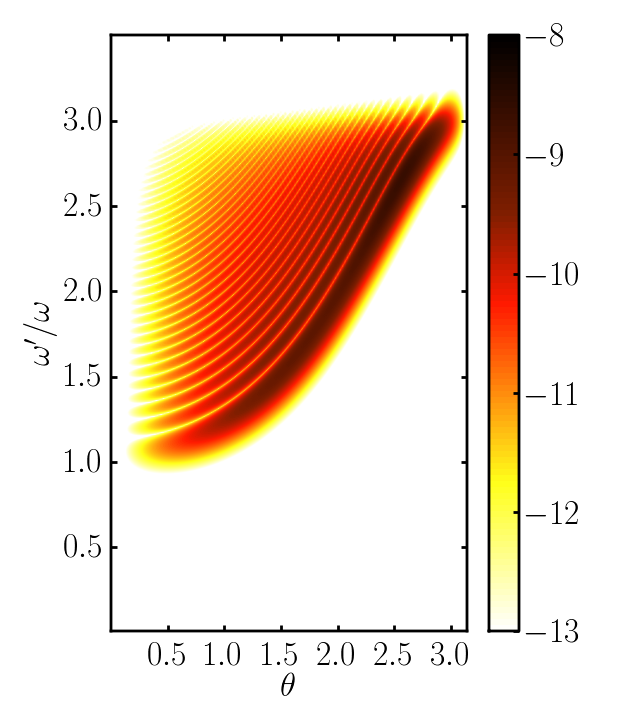}
\caption{The differential emission probability for the first three harmonics $\ell = 1,2$ and $3$ (from left to right panels) in the special reference frame as a function of frequency scaled frequency $\omega'/\omega$ and scattering angle $\theta$. The color code denotes the decadic logarithm of $dN_\ell / d\omega' d\Omega$ in units of inverse eV.
The laser has an intensity parameter of $a_0=2$ and a pulse length parameter of $\tau=20$.}
\label{fig.harmonics}
\end{figure}

One may define the partial differential photon emission probability for a single harmonic by
\begin{eqnarray}
 \frac{d N_\ell}{d\omega' d\Omega} 
  &=& \frac{1}{2} \sum_{\rm spin,pol} \frac{e^2 \omega'}{{64}\pi^3 k\cdot p k\cdot p'} \left| \mathrsfs M_\ell \right|^2, \label{eq.def.Nl}
\end{eqnarray}
such that
\begin{eqnarray}
  d N &=& \sum_\ell dN_\ell + \rm interferences.
  \label{eq.interferences}
\end{eqnarray}
The partial differential emission probability $dN_\ell / d\omega' d\Omega$ is exhibited in
Fig.~\ref{fig.harmonics} for the first three harmonics $\ell = 1,2$ and $3$ in
the special reference frame as a function of scaled frequency $\omega' / \omega$
and scattering angle $\theta$.
The spectra are shown for $a_0=2$ and a pulse length of $\tau = 20$ for a hyperbolic secant pulse shape.
In general, the interferences between different harmonics are important for differential observables, in particular, for energy resolved
spectra, for not too high values of $a_0\sim1$.
There, the substructures of the harmonics, which can be seen in Fig.~\ref{fig.harmonics},
yield interesting spectral information on the non-linear Compton scattering in short laser pulses
\cite{Seipt:PRA2011,Mackenroth:PRA2011}.
However, for high laser strength, $a_0 \gg 1$, the differential photon
emission probability (\ref{eq.def.Nl}) is a rapidly oscillating function of the photon energy $\omega'$.
The number of peaks per harmonic is given by $ |\Delta f | / \pi \propto |\beta \tau|$ and
grows, according to (\ref{eq.def.beta}), $\propto a_0^2$.
That means,
when measuring the energy spectrum with a spectrometer with finite energy resolution,
one actually obtains
an averaged spectrum. Denoting this average by $\langle \ldots \rangle $, we note that
\begin{eqnarray}
 \langle dN  \rangle  &=& \sum_\ell \langle dN_\ell \rangle 
\end{eqnarray}
which means that the interference terms in (\ref{eq.interferences}) average to zero.
Also the total photon yield $N$ is determined by the diagonal elements alone, e.g.~$\displaystyle N = \sum_\ell N_\ell$, i.e.~the off-diagonal elements give no contribution in the case of
$a_0\gg1$.
To calculate the photon yield, it is necessary to evaluate the coefficients (\ref{eq.def.c}). One could solve this problem by a direct numerical
integration. This works very well for not too large values of $a_0$ and $\beta$.
For higher laser intensity, e.g.~already for $a_0 > 20$, $\beta$ can become very large such that the phase exponential is in fact a rapidly oscillating function making a direct numerical integration very difficult.

For rapidly oscillating phase integrals,
the stationary phase technique can be applied to the integrals (\ref{eq.def.c}) as done e.g.~\cite{Narozhnyi:JETP83}. However, even for large $a_0$ there are regions in phase space where the stationary phase method is inapplicable:
\begin{enumerate}
 \item[1.)] In the vicinity of the non-linear monochromatic resonance at $s-\ell = \beta$, the stationary points are located at the center of the pulse and very close to each other. There, the
  the first derivative $g'(\phi_\star)$ is almost zero and, therefore, the stationary phase approximation tends to diverge. The exponent has to be expanded up to the third order derivative~\cite{Narozhnyi:JETP83} of the phase.
 \item[2.)] The stationary phase approximation is appropriate for large $|\beta| \gg 1$ only. However, in the vicinity of the forward scattering direction
    $\theta=\pi - \vartheta$ it behaves as
    $|\beta| =  a_0^2 \frac{\omega'}{\omega} \frac{\vartheta^2}{2}$, thus,
    the stationary phase method can be applied for angles $\vartheta^2 \gg a_0^{-2} \frac{\omega}{\omega'}$ only.
	Therefore, in the phase space regions where $|\beta|$ is small, a direct numerical evaluation
	may be used.
\end{enumerate}

Fortunately, these evaluation techniques complement one another, such that they may be combined together by suitable matching conditions $\mathfrak P_{1,2}$ (1 and 2 refer to conditions 1.) and 2.) above) to allow for an accurate calculation of the non-linear Compton scattering spectra.
The choice of these parameters is motivated in Appendix \ref{app.spa}.
(We note that
this method is not restricted to the non-linear Compton scattering process. It can be easily transferred to other strong-field processes such as stimulated pair production~\cite{Heinzl:PLB2010} or one-photon annihilation of $e^+e^-$ pairs~\cite{Ilderton:PRA2011}.)

\begin{figure}[!t]
  \includegraphics[width = 8cm]{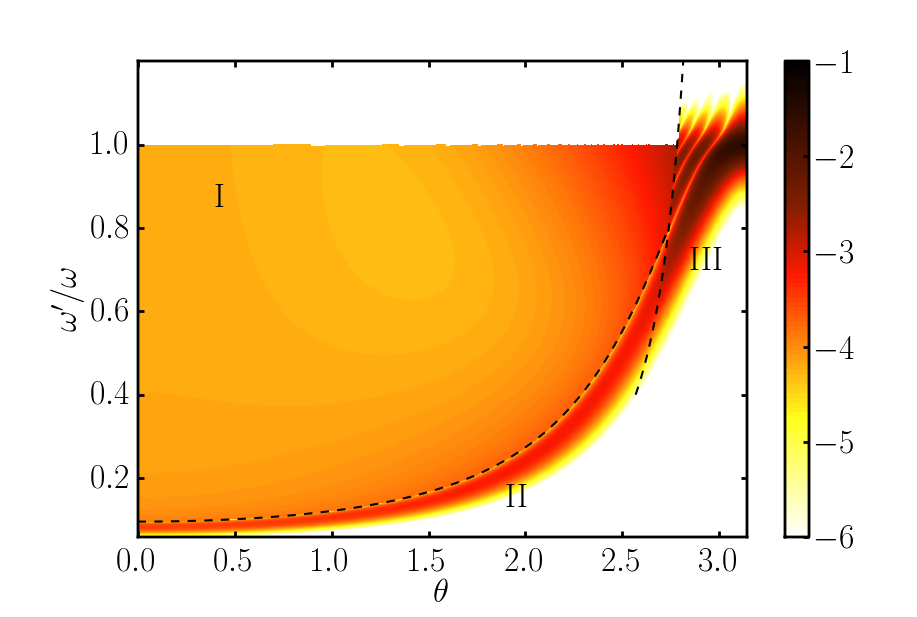} 
 \caption{ Differential photon yield $dN_1/d\omega'd\Omega$ as a function of
       scaled frequency $\omega'/\omega$ and scattering angle $\theta$ for $a_0=5$, $\tau=25$,  and a hyperbolic secant pulse envelope
       in the special reference frame.
       In the different areas of phase space (I, II and II, which are delineated by dashed curves),
       different methods of evaluation for the coefficients
       $\mathfrak c^m_n$ are employed.
       Parameters are $\mathfrak P_1 = 2.33$ and $\mathfrak P_2 = 10$.
}
 \label{fig.regions}
\end{figure}

We end this section by giving the formula for the phase space averaged photon emission probability which reads
\begin{widetext}
\begin{eqnarray}
  \left\langle \frac{d^2N_\ell}{d\omega'd\Omega} \right\rangle
  &=&\frac{e^2 \omega'}{64 \pi^3 k\cdot p k\cdot p'} 
      \Bigg\{ (4p\cdot p' - 8m^2) (\mathfrak c^0_\ell)^2 - 2 m^2 a_0^2 \, \mathfrak c^0_\ell \mathfrak c^{2}_\ell
           \nonumber \\ & & 
      + \frac{m^2a_0^2}{2} 
	  \left(
	    \frac{k\cdot p}{k\cdot p'} + \frac{k\cdot p'}{k\cdot p}
	  \right) [(\mathfrak c^1_{\ell+1})^2+(\mathfrak c^1_{\ell-1})^2]    
         -2  \alpha {k\cdot k'}
	  \,  \mathfrak c^0_\ell ( \mathfrak c^{1}_{\ell+1} +   \mathfrak c^{1}_{\ell-1} )
      \Bigg\} \label{eq.averaged.N}
\end{eqnarray}
\end{widetext}
with the averaged coefficients $\mathfrak c^m_n$
\begin{eqnarray}
 \mathfrak c_{\ell-n}^m &=&  
\left\{
\begin{array}{lll}
J_{\ell-n}( \alpha \sqrt{(s-\ell)/\beta}) \sqrt{\frac{s-\ell}{\beta}}^m 
    \sqrt{ \frac{2\pi}{\Big|\beta \sqrt{\frac{s-\ell}{\beta}} g'(\phi_\star)\Big|}} 
      & y \geq \mathfrak P_1 \text{ and } |\tau\beta| \geq \mathfrak P_2 & \hat \equiv \ {\rm I}, \\
\displaystyle 
 2\pi J_{\ell-n}( \alpha ) \frac{1}{|\beta g''(0)|^{\frac{1}{3}} }
	{\rm Ai}(-y)
      & y < \mathfrak P_1 \text{ and } |\tau\beta| \geq \mathfrak P_2 & \hat \equiv\  {\rm II},\\
      \displaystyle
 \int_{-\infty}^\infty d\phi J_{\ell-n}(\alpha g) g^m(\phi) e^{i(s-\ell)\phi - i\llangle f\rrangle} 
      & |\tau\beta| < \mathfrak P_2 &\hat \equiv \ {\rm III}
\end{array}
\right.
\end{eqnarray}
with $y=(s-\ell -\beta)/|\beta g''(0)|^{\frac{1}{3}}$.
In Appendix~\ref{app.spa} we provide further details on the evaluation of the stationary phase approximation.
The regions I, II and III, where the different evaluation schemes for the coefficients apply,
are exhibited in
Fig.~\ref{fig.regions}.
The displayed pattern persists for large values of $a_0 \gg 1$.

\section{Numerical Results}
\label{sect.numerical}

\begin{figure}
\includegraphics[width = 8cm]{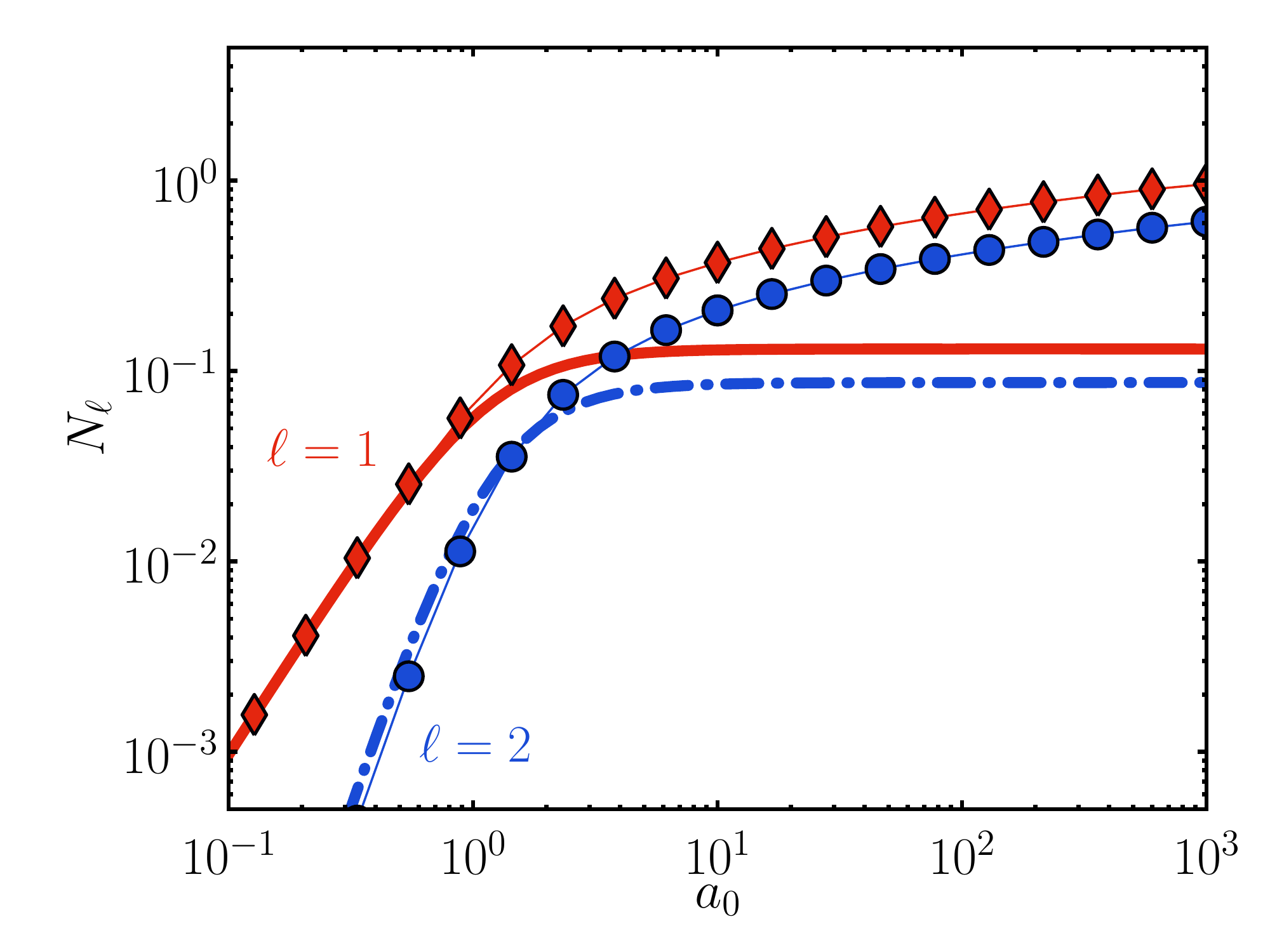}
\includegraphics[width = 8cm]{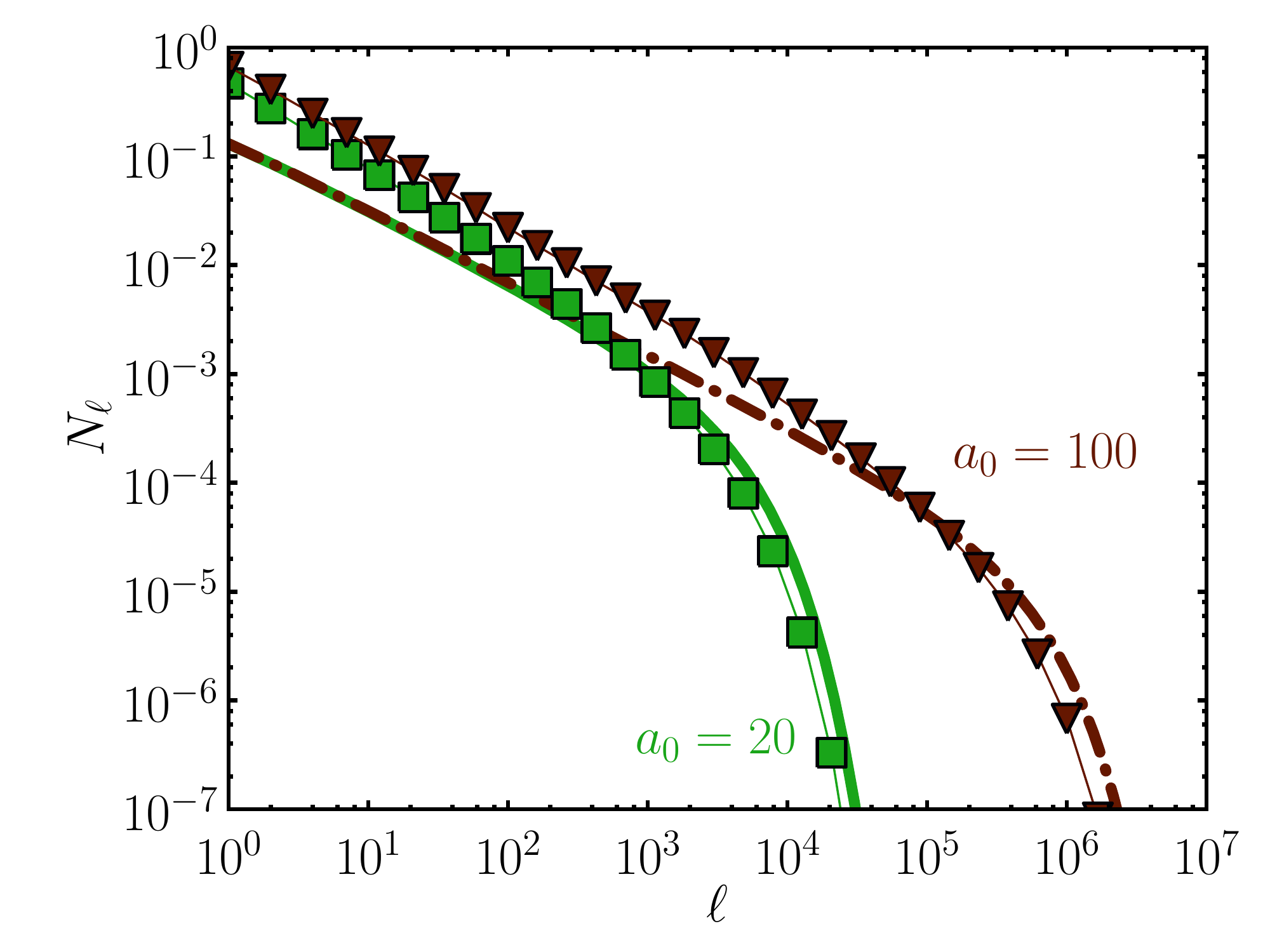}
\caption{Partial photon yield in hyperbolic secant pulse with $\tau=20$ (symbols) compared to the appropriate yield in a monochromatic wave (thick solid and dashed curves).
Matching parameters are $\mathfrak P_1=2.33$ and $\mathfrak P_2 = 20$.
Left panel: Behavior of the first two harmonics $N_1$ (red diamonds and solid curve) and
$N_2$ (blue circles and dash-dotted curve) as a function of $a_0$.
Right panel: Partial emission probabilities $N_\ell$ for different laser strength $a_0=20$ (green squares and solid curve) and $a_0=100$ (brown triangles and dashed curve).
}
\label{fig.cross.section} 
\end{figure}

For the numerical evaluation, we consider a laser with frequency (in the laboratory system) of $\omega = 1.5 \, \rm eV$ colliding head-on with
an electron beam with energy of $40 \, \rm MeV$. All calculations are performed in the special reference frame.
The partial photon yield $N_\ell$  for the first and second harmonic per electron in a hyperbolic secant pulse of duration $\tau=20$
are presented in Fig.~\ref{fig.cross.section},
where it is compared to the appropriate probability expected from a monochromatic pulse.
To get a finite number for the photon yield in a monochromatic field comparable with the pulsed laser
field we multiply the photon rate $\dot N_\ell$ in a monochromatic wave with the effective interaction time ${T_{\rm eff}}$ (see eq.~(\ref{eq.Teff})),
$N_\ell = {T_{\rm eff}} \dot N_\ell$, which is a characteristic number for each pulse shape.
This corresponds to a normalization to the same energy contained in the laser pulse.
Within this approach, both the pulsed and monochromatic photon yields coincide for low $a_0 \ll 1$,
i.e.~in the linear interaction regime, the dependence on the pulse shape drops out.
For large values of $a_0$, the photon yield in a pulsed laser field is enhanced by almost
a factor of ten as compared to the photon yield in a monochromatic wave.
The differences between pulsed and monochromatic yields at large values of $a_0$ express non-linear finite-size effects.

In the right panel of Fig.~\ref{fig.cross.section}, we display the partial photon yields for fixed $a_0=20$ and $100$ as a function of the harmonic number $\ell$ including 
harmonics up to $\ell=10^7$. The cutoff harmonic in a monochromatic laser field can be estimated from
the behavior of the Bessel functions at high index and argument and is given by $\ell_{\rm max} \sim a_0^3$.
While for low harmonics the photon yield $N_\ell$ is larger in pulsed fields, the ordering of the
two curves changes for high harmonics, where the photon yield in a pulsed field pulsed fields is smaller than in a monochromatic wave.

\begin{figure}[!t]
  \includegraphics[width = 8cm]{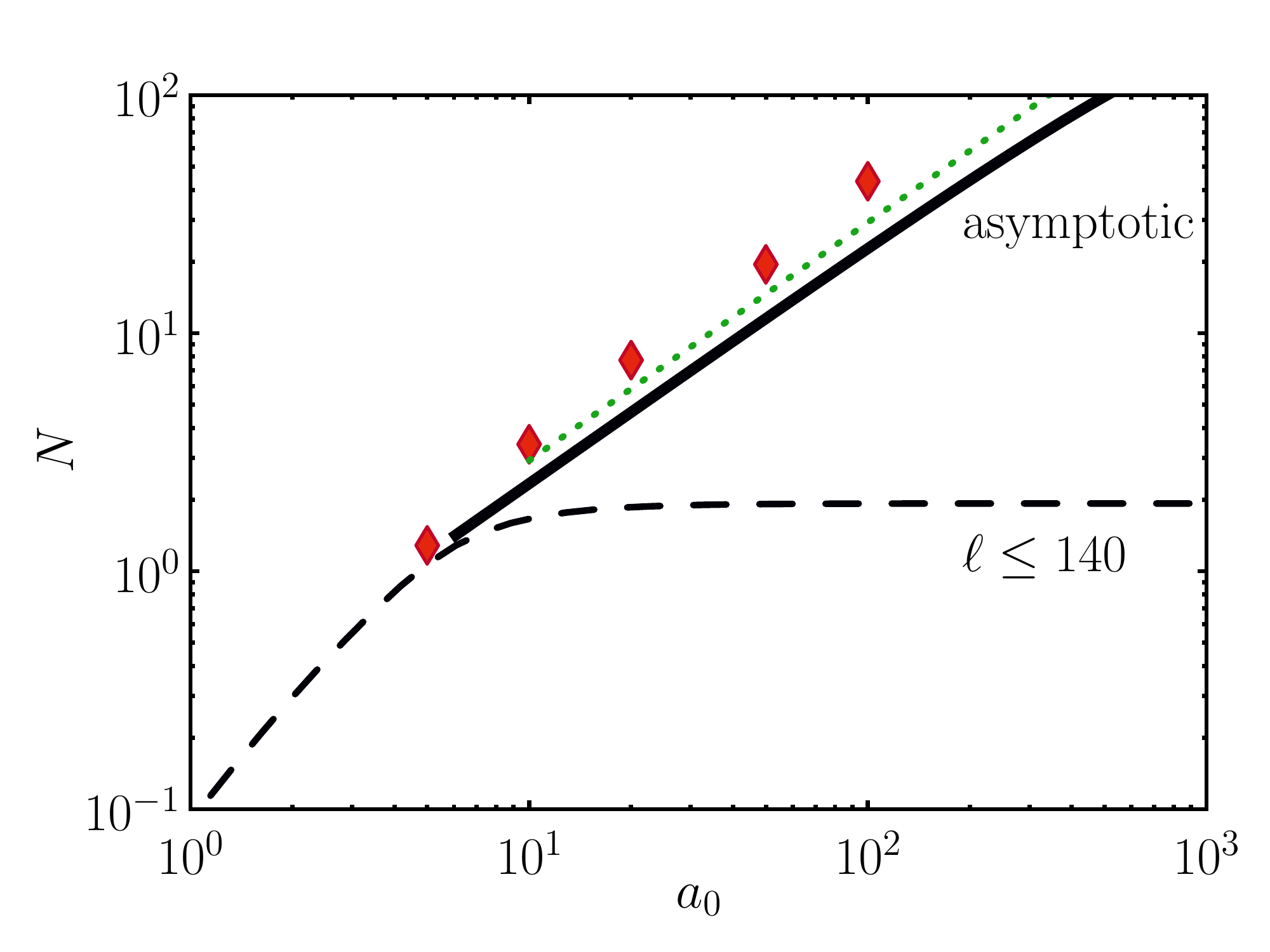} 
 \caption{ The total photon yield $N$ as a function of $a_0$. The result for pulsed laser fields is shown
 by symbols
(red diamonds) being a factor of two above the asymptotic monochromatic result (black solid curve, cf.~\cite{Titov:PRD2011}). The black dashed curve shows the sum of the first 140 harmonics in a monochromatic wave (compare \cite{Titov:PRD2011}). The green dotted curve is a
rough approximation to the asymptotic monochromatic result (see the text).
 }
 \label{fig.asymptotic}
\end{figure}

Comparing the total photon yields $N = \sum_\ell N_\ell$, summed over all harmonics up to
$\ell = 10^7$ in Fig.~\ref{fig.asymptotic}, we find that the total photon yield in a pulsed field exceeds the monochromatic result by a factor of two for $a_0 = 100$.
For the monochromatic result we show the sum over the first $140$ harmonics (dashed curve)
and an asymptotic approximation (representing the limit $a_0\to \infty$) being equivalent to a constant crossed field (solid curve, see e.g.~\cite{Ritus:JSLR1985,Titov:PRD2011}).
Additionally, the dotted curve is a rough estimate for the asymptotic probability
$N = \alpha_{QED} a_0 \Delta \phi$ \cite{diPiazza:PRL2010}, where we take $\Delta \phi = \omega T_{\rm eff}$ as the
relevant phase interval, and $\alpha_{QED}$ is the fine structure constant.
For the hyperbolic secant pulse this yields $\Delta \phi = 2\tau$.

\section{Summary and Conclusion}
\label{sect.summary}

In summary, we calculate the photon yield in non-linear Compton scattering at ultra-high intensity
$a_0 \gg 1$ for pulsed laser fields.
We use different methods in different parts of phase space,
including a stationary phase approximation, where the non-linear phase exponent is very large, and a direct numerical integration in phase space regions
where this is not the case.
At ultrahigh intensity, an averaging over the substructures in the
energy spectrum is included.
We find significantly modified partial and total photon yields in
pulsed laser fields when comparing to the case of monochromatic laser fields.
The partial yield for the first harmonic can be up to a factor of ten larger than the
corresponding monochromatic result. For the highest relevant harmonics for a given $a_0$
on the order of $\ell \sim a_0^3$, the photon yield in pulsed fields is typically lower than in monochromatic laser fields.
Nonetheless, the summed total probability is by a factor of two larger than the monochromatic result
which is typically approximated by a constant crossed field at large values of $a_0$.
This shows that the constant crossed field approximation is not as good for pulsed laser fields.
The method presented here is also be applicable to other processes like stimulated pair production~\cite{Heinzl:PLB2010}
and one-photon annihilation~\cite{Ilderton:PRA2011}.

Finally, we discuss the relevance of radiation reaction.
The radiation reaction is relevant if the parameter $R = \alpha_{QED} \chi a_0$ reaches or exceeds unity, where $\chi = a_0 k \cdot p / m^2$ is the quantum nonlinearity parameter.
For the parameters used for our numerical calculations and $a_0 = 100$ one finds $R = 0.04$,
i.e.~radiation reaction effects can be neglected. For even higher values of $a_0$, however, radiation reaction effects become important. Furthermore, the fact that the quantity related to the emission probability exceeds unity
is a hint that multi-photon emission, which is related to
radiation reaction \cite{diPiazza:PRL2010}, becomes important at these intensities.

\section*{Acknowledgments}

The authors gratefully acknowledge enlightening discussions with K.~Ledingham, R.~Sauerbrey and A.~I.~Titov.

\appendix

\section{Beyond the slowly varying envelope approximation}
\label{app.beyond.svea}
In this section, we go beyond the slowly varying envelope approximation. We give a further justification of the approximation and
comment on carrier envelope phase effects. The oscillating part of the non-linear phase exponent has the form
\begin{eqnarray}
 \tilde f(\phi) 
    &=&  \int d\phi 
		    \Big[
			 \alpha_1 g \cos(\phi+\hat\phi) + \alpha_2 g \sin(\phi+\hat\phi) 
			  +\beta \cos2 \xi g^2 \cos 2(\phi + \hat\phi) 
                    \Big]
\end{eqnarray}
introducing the carrier envelope phase $\hat\phi$.
This leads to integrals of the form
\begin{eqnarray}
I_n = \intop_{-\infty}^\infty d\phi \Big[g(\phi) e^{i\phi} \Big]^n, \qquad n \in (1,2)
\end{eqnarray}
which can be evaluated for a hyperbolic secant pulse with the substitution $x = e^{\phi/\tau}$ as
\begin{eqnarray}
\tilde f_1 &=& \tau {\rm Re}
	\left\{ 
	    \frac{1}{a} (\alpha_1 - i\alpha_2) e^{i\hat\phi} x^{2a} \, _2F_1(1,a;a+1; -x^2 )
	\right\} , \label{eq.f1.beyond.svea}\\
\tilde f_2 &=& \beta \tau \cos2\xi \, {\rm Re}
	\left\{ 
	    \frac{1}{a} e^{2i\hat\phi} x^{4a}\, _2F_1(2,2a;2a+1; -x^2 )
	\right\} ,\label{eq.f2.beyond.svea}
\end{eqnarray}
with $\tilde f = \tilde f_1 + \tilde f_2$, $a = (i\tau + 1)/2$ and the hypergeometric functions $_2F_1$.
For short pulses $\tau \ll 1$, $a\to 1/2$ becomes real.
Then, the hypergeometric functions take the limiting values
\begin{align}
 _2F_1(1,a;a+1; -x^2 )     & \to \  _2F_1(1,\frac{1}{2};\frac{3}{2}; -x^2 ) = \arctan x,\\
 _2F_1(2,2a;2a+1; -x^2 ) & \to \  _2F_1(2,1;2; -x^2 ) = \frac{1}{1+x^2},
\end{align}
such that
\begin{align}
\tilde f_1 &= 2 \tau (\alpha_1 \cos \hat \phi + \alpha_2 \sin \hat \phi ) \arctan \exp {\frac{\phi}{\tau}}, \\
\tilde f_2 &= \tau \beta \cos 2 \xi \, \cos 2 \hat \phi \, \tanh \frac{\phi}{\tau} .
\end{align}
Thus, the complete phase reads for $\tau \ll 1$
\begin{align}
f(\phi) &= \tau (\alpha_1 \cos \hat \phi + \alpha_2 \sin \hat \phi ) \arctan \exp{\frac{\phi}{\tau}}
+ \tau \beta  (\cos^2\xi \, \cos^2 \hat \phi +\sin^2 \xi \sin^2\hat \phi ) \tanh \frac{\phi}{\tau},
\end{align}
recovering the result of \cite{Mackenroth:PRA2011} for a single-cycle laser pulse with linear polarization, $\xi=0$, and $\hat \phi=0$.

In the opposite limit of long pulses, $\tau \gg 1$, $a$ becomes purely imaginary, $a \to i\tau/2$.
The slowly varying envelope approximation can be obtained from eqs.~(\ref{eq.f1.beyond.svea}) and (\ref{eq.f2.beyond.svea}) by approximating
$_2F_1(n,na;na+1;-x^2) \to \, _2F_1(n,na;na;-x^2) = (x^2 + 1)^{-n}$ yielding eventually Eq.~(\ref{eq.f.tilde.svea}).
Thus, our general result eqs.~(\ref{eq.f1.beyond.svea}) and (\ref{eq.f2.beyond.svea}) contains both
the previous results in the slowly varying envelope approximation for $\tau \gg 1$
and the single-cycle laser pulses for $\tau \ll 1$.

\begin{figure}[!t]
\includegraphics[width = 8cm]{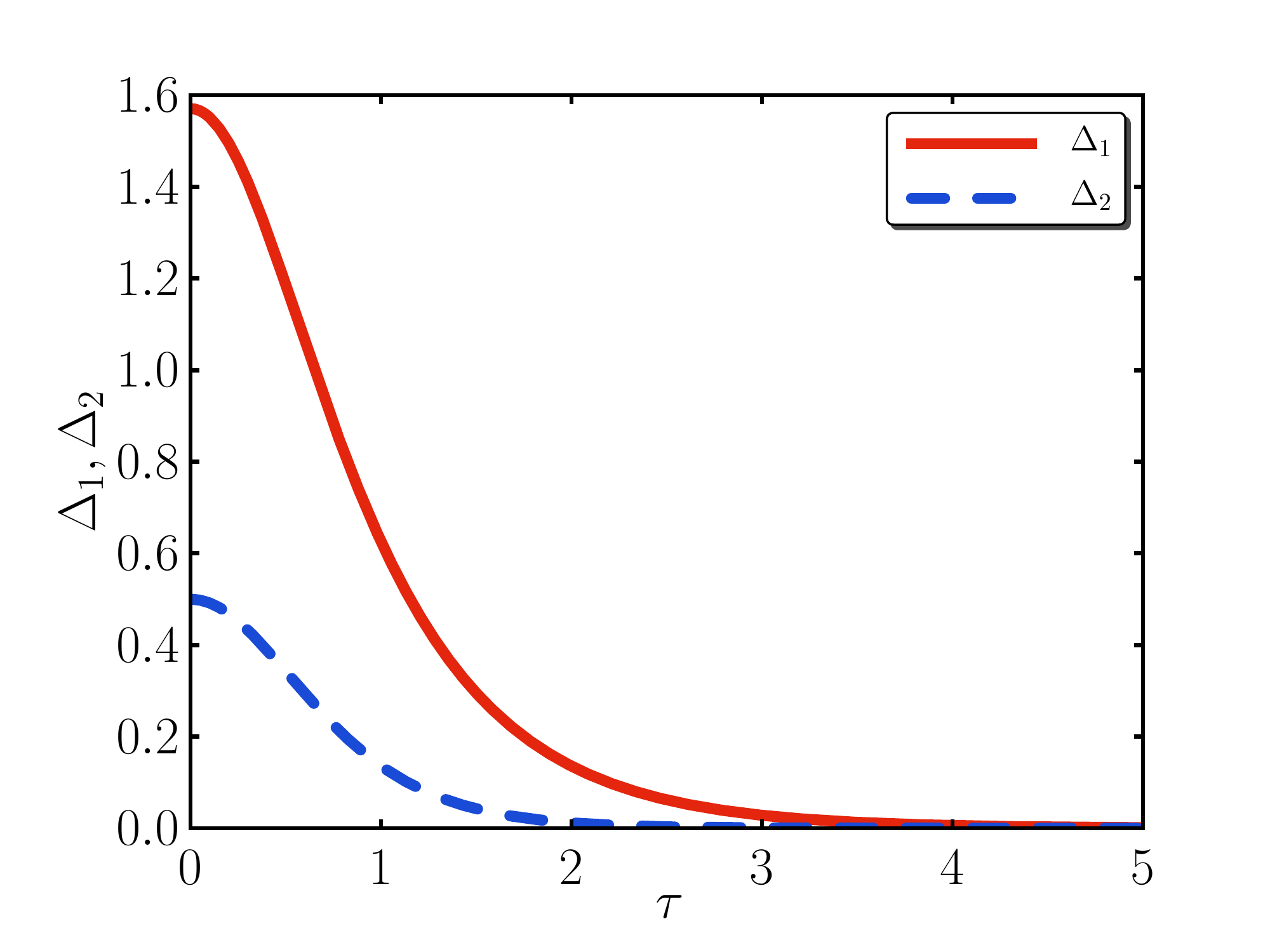}
\caption{The phase shift functions $\Delta_1$ and $\Delta_2$ as a function of pulse length $\tau$.}
\label{fig.Deltas} 
\end{figure}

One of the main differences between the slowly varying envelope approximation
and the full result is the build-up of an additional
phase shift when going from the distant past, $\phi=-\infty$, to distant future, $\phi=+\infty$, in addition
to the ponderomotive phase shift $\Delta f = 2\beta\tau$.
In the slowly varying envelope approximation this phase shift is equal to zero.
In the exact expressions for the phase, however, the additional phase shift is non-zero and depends on the carrier envelope phase:
\begin{eqnarray}
 \Delta \tilde f_1 &=& 2 \tau  \left( \alpha_1 \cos\hat\phi + \alpha_2 \sin \hat\phi \right) \Delta_1 , \\
 \Delta \tilde f_2 &=& 4 \tau \beta \cos 2\xi \,   \cos 2\hat\phi \, \Delta_2 ,\\
 \Delta_1  &=& \frac{1}{2} {\rm Re} \left\{ \psi(\frac{a+1}{2}) - \psi(\frac{a}{2}) \right\},\\
 \Delta_2  &=& \frac{1}{2} {\rm Re} \left\{ 1 + (1-2a) [\psi(\frac{2a+1}{2}) - \psi(a)] \right\},
\end{eqnarray}
where $\psi(z) = \frac{d}{dz} \log \Gamma(z)$ is the digamma function. Thus, the total phase shift becomes
now $\Delta f = 2\beta \tau + \Delta \tilde f_1 + \Delta \tilde f_2$, where the first term is the ponderomotive phase shift 
originating from $\llangle f\rrangle$.

The functions $\Delta_1$ and $\Delta_2$, depending only on the pulse length $\tau$, are depicted in Fig.~\ref{fig.Deltas}.
The additional phase is relevant for $\tau < 5$ only. For longer pulsed the phase shifts drop to zero exponentially fast.
This explains why the slowly varying envelope is much better than expected; even down to $\tau=5$ although it appears as an expansion in inverse powers of $\tau$.
It should be clear that this additional phase shift modifies the positions
of the stationary phase points and, therefore, also the support and number and position of zeros of
the coefficient functions $c^m_{\ell-n}$.

\section{Stationary phase, zero convexity approximations and matching conditions}
\label{app.spa}
\subsection{Stationary phase approximation}

The stationary phase approximation of (\ref{eq.def.c}) at a certain value of $s$ is given by the
sum of the contributions from the
two
stationary points
\begin{eqnarray}
 c_{\ell-n}^m(s-\ell) &=&  
	    J_{\ell-n}\big(\alpha g(\phi_\star)\big) g^m(\phi_\star)  \sqrt{\frac{2\pi}{|\beta G_2''(\phi_\star)|}}
	    \exp \left\{i(s-\ell)\phi_\star - i \beta G_2(\phi_\star) \red{-} i\frac{\pi}{4} \right\} \nonumber \\
      & & + J_{\ell-n}\big(\alpha g(-\phi_\star)\big) g^m(-\phi_\star)  \sqrt{\frac{2\pi}{|\beta G_2''(-\phi_\star)|}}
	    \exp \left\{-i(s-\ell)\phi_\star - i \beta G_2(-\phi_\star) \red{+} i\frac{\pi}{4} \right\} \nonumber \\ 
  &=&   2 J_{\ell-n}\big(\alpha g(\phi_\star)\big) g^m(\phi_\star) \sqrt{\frac{2\pi}{|\beta G_2''(\phi_\star)|}} 
	     \cos \left\{ (s- \ell) \phi_\star -  \beta G_2(\phi_\star) \red{-}  \frac{\pi}{4} \right\}.
	     \label{eq.c.SPA}
\end{eqnarray}
The oscillations of the coefficient functions $c^m_{\ell-n}$ stem from the cosine term which is
a common factor in all contributions to the partial matrix element $\mathrsfs M_\ell$.
When squaring the matrix element to calculate the partial differential
photon emission probability $dN_\ell$, we obtain a $\cos^2$ behavior, which averages to $1/2$. Because the interferences cancel on average,
we may replace the cosine in (\ref{eq.c.SPA}) by $1/\sqrt2$ and write for the averaged function
\begin{eqnarray}
 \langle c_{\ell-n}^m(s-\ell) \rangle &=& J_{\ell-n} \big( \alpha g(\phi_\star) \big) g^m(\phi_\star) \sqrt{\frac{4\pi}{|\beta G_2^{\prime\prime}|}}.
\end{eqnarray}
We now estimate the number of zeros of $c^m_{n-\ell}$ which is also the number of
subpeaks in a given harmonic.
The functions $c^m_{n-\ell}$ are zero when
the contributions from the two symmetric stationary points interfere destructively;
i.e.~whenever the phase fulfills the equation
\begin{eqnarray}
 \beta \left[ \phi_k g^2(\phi_k) - \int_0^{\phi_k} d\phi g^2(\phi) \right] - \left(\frac{1}{4}-k\right)\pi = 0
 \label{eq.zeros}
\end{eqnarray}
defining a series $(s_k)_{k = 1\ldots K}$
of zeros of $c_{\ell-n}^m(s-\ell)$.
The number $K$ of zeros is determined by the range of values the term in the square brackets can take restricting the largest and smallest allowed values of $k$.
Noting that the function in the square brackets in (\ref{eq.zeros})
is a monochromatic dropping function of its argument
it suffices to consider its values at $\pm \infty$. However, the first term goes to zero when
approaching infinity and the second term (including the leading $\beta$ term) is $\llangle f\rrangle$
. Thus, $\pi (k_{\rm max} - k_{\rm min}) \sim |\llangle f\rrangle(\infty) - \llangle f\rrangle(-\infty)|$
and the number of zeros
$K$ is determined by the total ponderomotive phase shift
$\displaystyle \Delta f$ as $K \sim  |\Delta f| / \pi  \propto |\beta\tau|/\pi$.
This result is a generalization of the findings of \cite{Hartemann:PRL2010} within the classical
theory of Thomson scattering for hyperbolic cosine pulse shapes in the backscattering direction
to the quantum theory of Compton scattering for arbitrary pulse shapes and arbitrary scattering angles.

\subsection{Zero convexity approximation}

In the vicinity of $\phi=0$ the stationary phase approximation diverges.
One must expand the exponent up to the third derivative, around the point of zero convexity
(i.e.~the second derivative of the phase in (\ref{eq.def.c}) vanishes) yielding the approximation
\begin{eqnarray}
c_{\ell-n}^m(s-\ell) & \cong & 2\pi J_{\ell-n}(\alpha) \left( \frac{1}{|\beta g''(0)|} \right)^{\frac{1}{3}} {\rm Ai} (-y), \\
y	&=& \frac{s-\ell-\beta }{|\beta g''(0)|^{\frac{1}{3}}},
\end{eqnarray}
where ${\rm Ai}(y) = \frac{1}{2\pi} \int_{-\infty}^\infty dt \exp \{ iyt + it^3/3\}$ is the Airy function.

The height and shape of the main peak of the spectrum
are determined by the curvature of $g(\phi)$ at the center of the pulse, where the second derivative vanishes.
The main peak is the same for hyperbolic secant and Gaussian envelopes, because the second derivative at the origin is the same.
The only difference comes from the different behavior of the first derivative in the stationary phase approximation.
This determines the different shape of the spectra in the high-energy tail of each harmonic.

\subsection{Matching Conditions}

Here we motivate our choices for the matching parameters $\mathfrak P_{1,2}$.
The parameter $\mathfrak P_1$ determines the transition between the stationary phase approximation
and the zero convexivity approximation.
We define the matching point by $y = \mathfrak P_1$ 
which should be in the range $ 1.4 < \mathfrak P_1 < 2.3381$ because for smaller $\mathfrak P_1$ the two stationary phase points would
be too close to each other~\cite{Narozhnyi:JETP83} and for larger values the two approximations
start to differ too much.
We use the stationary phase approximation for $y \geq \mathfrak P_1$ (region I in Fig.~\ref{fig.regions})
and zero convexity approximation for $y < \mathfrak P_1$  (region II in Fig.~\ref{fig.regions}).

The parameter $\mathfrak P_2$ determines the matching between the direct numerical evaluation and the stationary phase/zero convexity approximation.
The approximations for the functions $c^m_{\ell-n}$ are used whenever
$|\tau\beta| > \mathfrak P_2$ and the full numerical results otherwise
(region II in Fig.~\ref{fig.regions}).
The numerical result was found to be rather insensitive to the explicit value in the range {$10<\mathfrak P_2<50$}.

%


\end{document}